\documentclass[aps,showpacs,prc,preprint,superscriptaddress,floatfix,nofootinbib]{revtex4}
\usepackage{amsfonts}
\usepackage{amsmath}
\usepackage{dsfont}
\usepackage{ulem}
\usepackage{xcolor}

\usepackage{amssymb}
\usepackage{amsfonts}
\usepackage{amsmath}
\usepackage{amssymb}
\usepackage{amsthm}
\usepackage{epsf}

\usepackage[dvips]{graphicx}
\newcommand{\boldm}[1] {\mathversion{bold}#1\mathversion{normal}}

\def\bk{{\mbox{\boldmath$k$}}}
\def\bq{{\mbox{\boldmath$q$}}}
\def\bp{{\mbox{\boldmath$p$}}}
\def\bgam{{\mbox{\boldmath$\gamma$}}}

\def\calP{{\cal P}}

\def\la{\langle}
\def\ra{\rangle}
\def\bk{  {\bf k}}
\def\bq{ {\bf q} }
\def\bp{ {\bf p} }
\def\bP{{\bf P}}
\def\be{\begin{eqnarray}}
\def\ee{\end{eqnarray}}
\begin{document}
\title{Pseudo-scalar {\boldm $     q\bar q  $} bound states at finite temperatures within a Dyson-Schwinger--Bethe-Salpeter approach }
 \author {S.~M. Dorkin}
\affiliation{Bogoliubov Lab.~Theor.~Phys., 141980, JINR, Dubna,  Russia}
\affiliation{International University Dubna, Dubna, Russia }
 \author{L.~P. Kaptari}
 \affiliation{Bogoliubov Lab.~Theor.~Phys., 141980, JINR, Dubna, Russia}
\affiliation{Helmholtz-Zentrum Dresden-Rossendorf, PF 510119, 01314 Dresden, Germany}
\author { B.~K\"ampfer}
\affiliation{Helmholtz-Zentrum Dresden-Rossendorf, PF 510119, 01314 Dresden, Germany}
\affiliation{Institut f\"ur Theoretische Physik, TU Dresden, 01062 Dresden, Germany}

\begin{abstract}
 The combined Dyson-Schwinger--Bethe-Salpeter equations are employed at non-zero temperature. The truncations refer to a rainbow-ladder approximation augmented with an interaction kernel which facilitates a special temperature dependence. At low temperatures, $T \to 0$, we recover a quark propagator from the Dyson-Schwinger (gap) equation which delivers, e.g. mass functions $B$, quark renormalization wave function $A$,
and two-quark condensate $\la  q \bar q \ra$ smoothly interpolating to the $T = 0$ results, despite the broken O(4) symmetry in the heat bath and discrete Matsubara frequencies. Besides the Matsubara frequency difference entering the interaction kernel, often a Debye screening mass term is introduced when extending the $T = 0$ kernel to  non-zero temperatures. At larger temperatures, however, we are forced to drop this Debye mass  in the infra-red part of the longitudinal interaction kernel to keep the melting of the two-quark condensate in a range consistent with lattice QCD results. Utilizing that quark propagator for the first few hundred fermion Matsubara frequencies we evaluate the Bethe-Salpeter vertex function in the pseudo-scalar $ q \bar q$ channel for the lowest boson Matsubara frequencies and find a competition of  $ q \bar q$ bound states and quasi-free two-quark
states at $T = {\cal O}$ (100 MeV). This indication of pseudo-scalar meson dissociation below the anticipated QCD deconfinement temperature calls for an improvement of the approach, which is based on an interaction adjusted to the meson spectrum at $T = 0$.
\end{abstract}

\maketitle

\section{Introduction}
 The description of mesons as quark-antiquark bound states within the framework of the Bethe-Salpeter (BS) equation
 with momentum dependent quark mass functions, determined by the Dyson-Schwinger (DS equation, is able
 to explain successfully many spectroscopic data, such as
 meson masses~\cite{fishJPA,Thomas,rob-1,MT,ourLast,Hilger,Maris:2003vk,Holl:2004fr,Blank:2011ha},
 electromagnetic properties
 of pseudoscalar mesons and their radial excitations~\cite{Jarecke:2002xd,Krassnigg:2004if,Roberts:2007jh}
  and  other   observables~\cite{ourFB,wilson,JM-1,JM-2,rob-2,Alkofer,fisher,Roberts:2007jh}.
 Contrary to purely  phenomenological models, like  the quark bag  model,
 such a formalism  maintains   important features of QCD, such as dynamical chiral
 symmetry breaking, dynamical quark dressing, requirements of the renormalization group theory etc., cf.~Ref.~\cite{physRep}.
 The main ingredients here are  the full quark-gluon vertex function and  the dressed gluon
 propagator,  which are entirely determined  by  the running coupling
 and the bare quark mass parameters. In principle, if one were able to solve the Dyson-Schwinger equations
 for the quark propagator, the gluon propagator, the quark-gluon and inter-gluon vertices ,
 supplemented by ghosts if necessary, then the approach would  not depend on  any additional parameters.
In practice,  one restricts oneself to calculations within effective models which
  specify the dressed vertex function $\Gamma_\mu$ (for quark-gluon coupling) and interaction kernel $D_{\mu\nu}$
  (for the gluon propagator).
  The rainbow-ladder approximation~\cite{MT} is a model with rainbow truncation of the
  vertex function $\Gamma_\mu\to \gamma_\mu$ in the quark DS equation  and
  a specification of the dressed quark-quark interaction kernel (for gluon exchanges)
   as $ g^2 D_{\mu\nu}(k) \to {\cal G}(k^2) D_{\mu\nu}^{free}(k)$ within the Landau gauge. (Here, $\gamma_\mu$
   is a Dirac gamma matrix and $D_{\mu\nu}$ stands for the gluon
   propagator; $g$ is the coupling strength and $k$ denotes a momentum.)

   The model is completely specified once a form
is chosen for the “effective coupling” ${\cal G}(k^2)$. The ultraviolet
behavior is chosen to be that of the QCD running coupling $\alpha_s(k^2)=g^2(k^2)/4\pi$ computed within one-loop
approximation; the ladder-rainbow truncation then generates the correct perturbative
QCD structure of the DS and BS
equations. Moreover, the  ladder-rainbow truncation preserves such an  important feature of the
theory as the maintenance of the Nambu-Goldstone theorem in the chiral limit,
according to which   the spontaneous chiral symmetry breaking results in an appearance of a
 (otherwise absent) scalar term in the quark propagator of the DS  equation. As a consequence, in the BS
 equation a massless pseudoscalar bound state should appear. By using the Ward  identities, it has been proven (see,
 e.g.  Refs.~\cite{smekal,scadron,munczek}) that, in the chiral limit, the DS equation for the quark propagator and
 the BS equation for a massless pseudo-scalar in ladder approximation are completely  equivalent.
 It  implies  that such a  massless bound state (pion) can be interpreted as a Goldstone boson.
 This results in a straightforward understanding    of the pion as  a Goldstone boson and  quark-antiquark
 bound state at the same time.

    Another important property of the DS and BS equations is that, at zero temperatures, or equivalently - in vacuum -
    they are explicitly Poincar\'{e}  invariant.    This  frame-independency of the approach provides
    a useful tool in studying processes where  a rest frame for mesons cannot or  needs not be defined.
    A rather different situation is encountered at finite temperatures,   where a  particle is embedded in
    a heat bath. In this case,  the propagator   of the particle is defined by  two four-vectors,
    the four-momentum $p$  of the particle and a unit four-vector $u$
    characterizing the heat bath itself~\cite{hatsuda,Ayale,RobProg}.
    The scalar parts of the propagator then depend on $p^2$ and $(up)$. An existence of the
    statistical thermal factor  $\exp(-  H/T)$, $H$ being the   QCD hamiltonian, implies the choice of the unit vector
    $u$  as $u=({\bf 0},1)$,   i.e. the heat bath at rest~\cite{hatsuda}.
 As a consequence, the O(4) symmetry of equations is lost and only the spatial translation invariance
     is maintained. This requires a separate treatment of the transversal and longitudinal, with respect to the heat bath,
   parts of the kernel with the need of additional functions in parametrizing  the quark propagators and the   BS vertex functions.

    In quantum field theory, a system at finite temperature can be
 described within the  imaginary-time formalism,
which uses the Matsubara frequancies~\cite{matzubara,kapusta,abrikosov}. Due to finiteness of the heat bath
 temperature $T$ and the requirements of periodicity (bosons) or antiperiodicity (fermions) of the
 considered  fields~\cite{periodic,KMS,kapusta,abrikosov},
  the Fourier transform  to  Euclidean momentum space becomes discrete in the time direction, resulting
 in a discrete spectrum of the energy, known as the Matsubara frequencies. Consequently, the
 interaction kernel and the DS and BS solutions must become also discrete with respect to these frequencies.
It is worth mentioning that there is no straight forward connection of the obtained solutions
in the discrete Euclidean space to  the corresponding continuum quantities in  Minkowski space~\cite{fisher1,roberts,yaponetz}.
In particular, within the Matsubara formalism one can define the mass of the bound state
in different ways in dependence of the choice of the total momentum $P=(\bP,i\Omega_N)$;
at $N=0$ one defines the so-called screening mass, which has no direct relation with the usual
inertial mass. At $|\bP|\ne 0$ and $N\ne 0$ the corresponding pole mass $P^2=-M^2$ is also discrete
and hitherto it is note clearly established  how to relate it to the inertial mass. In principle,
at each particular choice  of $P$
one can define a quark correlator and try to define the inertial mass by computing the
momentum Fourier transforms and the resulting spatial integral  of the correlator (analogue of the 2-points  Schwinger
functions, see e.g.~\cite{robertsSchwinger}) and   by taking the time derivative of the logarithm
 of the obtained expression, see also~\cite{fisher1,roberts,yaponetz,physrepMorley,blaske}.
 Evidently, such a procedure requires knowledge of the pole mass for all, or at least for a large enough number
 of Matsubara frequencies. It is worth mentioning that a  growing interest nowadays is in approaches based on
 real-time calculations of quark and gluon propagators and their spectral functions within the functional
 renormalization group~\cite{frg1,frg2, frg3}. These approaches also require analytical continuation of
 the propagators from Euclidean to Minkowski spaces.

 In the present paper, we calculate the pole masses of pseudo-scalar
 quark-antiquark bound-states only for the few
 first Matsubara frequencies and investigate their dependence on temperature in a large interval, including
 the "dissociation region". We anticipate that the main peculiarities
 of  masses computed  at some particular Matsubara frequencies will
 reflect the general $T$ dependence of the inertial masses at high temperatures.
We treat the bound states within the BS formalism within the
same approach as the one used in solving the DS equation, i.e. with the rainbow truncation and
AWW interaction kernel~\cite{Alkofer}.
Recall that the merit of the approach is that, once the effective parameters are fixed
 (usually the effective parameters of the
 kernel are chosen, cf. Refs.~\cite{Parameterslattice,williams},  to reproduce the known
 "experimental" data from lattice calculations, such as the quark mass function and/or  quark condensate,
 the whole spectrum of known mesons is supposed to  be described,   on the same footing,
 including also excited states. The achieved  amazingly good description  of
 the mass spectrum with only  few effective parameters encourages one to employ the same approximations
 to the  DS and BS equations also at finite temperatures with the hope that, once an adequate
 description of the quark propagators at non-zero temperature    is accomplished, the corresponding solution
 can be implemented into the BS equation for mesons  to investigate the meson properties
 in hot matter.

   At low temperatures, the properties of hadrons   in nuclear matter
   are expected to change in comparison with the vacuum ones, however, the main  quantum
   numbers, such as  spin and orbital momenta, space and inner parities etc. are maintained.
   The hot environment may modify the hadron masses, life times  (decay constants) etc. Contrarily,
   at sufficiently large temperature in hot and dense strongly interacting matter,
     (phase) transitions may occur, related to quark deconfinement phenomena, as e.g.
   dissociation of hadrons into quark-gluon degrees of freedom. Therefore,
   this temperature region is of great interest,
   both from a theoretical and experimental point of view. So far,
   the truncated DS and BS formalism has  been mostly used
    at large   temperatures  to investigate the critical phenomena
    near and above the  pseudo-critical and  (phase) transition values predicted by lattice simulation data
   (cf. Refs.~\cite{kitaizyRpberts,MTrenorm,rischke,ourModernPhys}).
  It has been found that, in order to achieve an agreement of   the  model results with lattice data,
  a modification of the vacuum interaction kernel is required.
  First of all it has been observed that modifications  of the
  (pure phenomenological,  nonperturbative) infra-red (IR) term by including the $T$ dependence of the
  perturbative Debye mass results in an essential suppression of the longitudinal interaction kernel
  at large temperatures~\cite{BlankKrass,ourModernPhys}.
  As a consequence, the predicted critical temperatures $T_c$ of chiral symmetry restoration
  in the chiral limit and also the critical temperatures for  the maximum of chiral susceptibility, as well as of
   inflection point of the mass function or of the quark condensate,
   are much smaller, by about 50\%, than that expected from the lattice calculations.
   It became evident that  modifications of the IR term at large temperatures are necessary.
   Namely, the IR  term has to vanish abruptly in this region and to be replaced by another
   phenomenological kernel.  For instance, it has been  suggested~\cite{kitaizyRpberts,robLast}  to employ
    a kernel with a  Heavyside step-like behaviour  in the vicinity
   of the (pseudo-) critical temperature  $T_c$. Then,  it becomes possible to  achieve a
  rather reliable description of such quantities as the quark spectral function,
  plasmino modes, thermal masses etc., see also Ref.~\cite{kitazavaPRD80}. However, a use   of  such a
   discontinuously  modified interaction in
  the BS equation in the whole temperature range becomes hindered.

  Another strategy of solving the DS equation in a
  larger interval of temperatures  is to  utilize directly   the available  results of nonperturbative lattice QCD (LQCD)
  calculations  to fit, point by point, the interaction kernel
  at given  temperatures. In such a way one achieves a good description
  of the quark mass function and condensate  for  different   temperatures,  including the region
  beyond  $T_c$~\cite{fischerPRD90,FischerRenorm}. The success of such approaches demonstrates that the
  rainbow approximation to the DS equation with a proper choice of the interaction kernel  is quite
  adequate in  understanding the properties  of quarks in  a hot environment. Nevertheless, for
   systematic studies of quarks
  and hadrons within the BS equation, on needs a  smooth parametrization of the
   kernel  in the whole interval of the considered
   temperatures. In view of still scarce LQCD data, such a direct parametrization from "experimental" data is problematic.
   An alternative method  is to solve simultaneously   a (truncated) set of Dyson-Schwinger equations
  for the quark and gluon propagators within some additional approximations~\cite{fischerGluon}.
  This approach also provides
  good description of the quark mass function and condensate   in vicinity of $T_c$,
  however it becomes too cumbersome in attempts
  to solve the BS equation,  since in this case one should solve a too large system of equations.
It should also be  noted that  there are other  investigations of the quark propagator
within the rainbow truncated   DS equation,
which   employ solely the vacuum parameters in  calculations of  $T$ dependencies of quarks~\cite{BlankKrass}
   without further attempts to accommodate    to LQCD results in the kernel. As a result    one
  finds that  the  critical behaviour of the propagators (e.g.  chiral symmetry restoration)  starts at temperatures
   much smaller than the ones  expected from  LQCD.

 In the present paper we are interested in a qualitative study of the masses of quark-antiquark bound states at moderate and
 large temperatures. Particular attention is paid to the region where the bound state mass becomes larger
 than the sum of two masses of free quarks. This temperature range probably indicates a dissociation of the
 ground bound state and existence of only highly excited correlations.
 For such a qualitative analysis we employ the simplest
 version of the kernel which consists of only the infra-red term of the Maris-Tandy~\cite{MT} kernel, known also as the
 Alkofer-Watson-Weigel~\cite{Alkofer}  kernel, referred  to as the  AWW model.
 For further kernels cf.~\cite{MT,robLast,BlankKrass,otherKernels1,otherKernels2,otherKernels3},
 in particular to symmetry preserving issues and going beyond the rainbow-ladder approximation.
  As mentioned above, including of the Debye mass in this term results in  too low values of the
  critical temperatures, and a proper modification  is needed. For this sake, in the present
  paper we ignore  the Debye mass which allows to obtain
  solutions of the DS equation with characteristics close to that of lattice calculations.
   However, for completeness we also
  present some results of calculations with accounting of the Debye mass too.
  We start with the AWW interaction kernel, known at $T=0$  to provide  a good description  of
 the mass spectrum of scalar, pseudo-scalar, vector and tensor mesons.
Then we extend the DS and BS equations  to finite temperatures  and solve the corresponding equations
in a large interval of $T$ and Matsubara frequencies. Results of calculations are compared with masses of free quarks.

 Our paper is organized as follows.
 In Sec.~\ref{s:bse},  we recall  the truncated Dyson-Schwinger
 (tDS) and truncated Bethe-Salpeter (tBS) equations in vacuum and at finite temperatures.
  The rainbow approximation for the DS equation kernel in vacuum is  introduced
  and  the  system of equations for the quark propagator is solved.

 Numerical solutions for the quark propagator at finite bare quark masses
 are discussed in Sec.~\ref{tDS}, where
 the inflection points of the quark condensate  and the mass function are considered as a
 definition of the pseudo-critical
 temperature $T_c$. It is shown that, for finite quark masses,
 the inflection method determines the pseudo critical temperatures by $\sim 50\%$ smaller than the ones
 obtained by other approaches, e.g. by LQCD calculations.
  It is found that, by merely ignoring  the Debye mass in the infra-red term, it becomes possible to
 reconcile the model and lattice QCD results; the such obtained values of
 $T_c\sim 135$~MeV is  only by about  $10-15$\% smaller then the lattice results (reported, e.g. in
 Refs.~\cite{yoki,FischerRenorm}).
 In Sec.~\ref{BSE} we generalize the tBS equation to  finite temperature by exploring   the same procedure  as
 that used in  generalizing the tDS equation to finite $T$. As already mentioned, due to the breaking of the O(4) symmetry,
 the tBS equation cannot be longer considered for mesons at rest. The total momentum $P^2 = -M^2_{qq}$ is replaced  as
 $P({\bf 0},iM_{qq}) \rightarrow P({\bf P},iE)$, where now the energy $E$ becomes discrete and depends on the Matsubara
 frequency $\Omega_N = 2\pi N T$.  In dependence of the chosen components of the total momentum $P$, one can
 define at least three different ground state  masses:
 (i) the thermal mass at ${\bf P}=0$, $M^2= (2\pi N)^2T^2$,
 (ii) the imaginary screening mass at zero Matsubara frequency $n=0$, ${\bf P}^2 = -M^2$ and
 (iii) the pole mass $M^2 = (2\pi N)^2T^2 - {\bf P}^2$. In principle, these masses are
 not obligatorily the same at given temperature $T$, as discussed in Refs.~\cite{fisher1, yaponetz,blaske,roberts}, for instance.
 In the present paper, we consider the pole masses which at
  given temperature and Matsubara frequency are continuous functions of the three momentum ${\bf P}^2$.

   Summary and conclusions are collected in Sec.~\ref{summary}.
\section{Basic Formulae}
\label{s:bse}
\subsection{Truncated Dyson-Schwinger  and Bethe-Salpeter    equations in vacuum}
\label{Bet}
To determine the bound-state mass of a quark-antiquark pair one needs to solve the
DS and the {homogeneous}   BS equations, which
in the rainbow ladder approximation    and in Euclidean space  read
\begin{eqnarray}&&
S^{-1}(p)= S_0^{-1}(p) + \frac 43 \int \frac
{d^4 k }{(2\pi)^4} \left[g^2 {D}_{\mu \nu}(p-k) \right]\gamma_{\mu} S(k)
\gamma_{\nu}\: ,
\label{sde}\\ &&
    \Gamma(P,p) =   -\frac 43  \int \frac {d^4k}{(2\pi)^4}
    \gamma_{\mu} S(\eta_1 P+k) \Gamma(P,k) S(-\eta_2 P+k))\gamma_{\nu}
    \left [g^2    { D}_{\mu \nu}(p-k)\right ] \: ,
\label{bse}
\end{eqnarray}
where $\eta_1$ and $\eta_2=1-\eta_1$ are the partitioning parameters  defining the quark momenta
as $p_{1,2}=k \pm \eta_{1,2} P $ with $P$, $p$ and $k$ denoting the total and relative momenta
of the bound system, respectively;\footnote{
  {Usually,  for quarks of masses $m_{1,2}$ the partitioning
parameters  are chosen as  $\eta_{1,2}=m_{1,2}/({m_1+m_2})$. However, in general, the tBS solution
is independent of the choice of $\eta_{1,2}$.    }}
 $\Gamma(P,k)$ stands for  the tBS vertex function being a $4\times 4$ matrix,
$S_0(p)=\left (i\gamma\cdot p +m\right)^{-1}$  and
$S(p)=\left ( i\gamma\cdot p A(p)  +B(p)\right)^{-1}$ are the
   propagators of bare and dressed  quarks, respectively, with mass parameter $m$ and
    the dressing functions  $A(p)$ and $B(p)$.  At zero temperature the above equations are O(4) invariant and
    the propagator functions $A(p)$ and $B(p)$  depend solely on $p^2={\bf p}^2+p_4^2$. The
    total momentum $P = ({\bf 0}, iM_{q\bar q})$, for a particle at rest,  is   an external parameter
    for~(\ref{bse}); the momenta of individual quarks are
    $p_{1,2}^2 = -M^2_{q\bar q}/4 +k^2 \pm iM_{qq} k_4$, where the Euclidean components
    $\bf k$ and $k_4$ of the relative momentum are defined as
     $k^2 = \bk^2 + k_4^2$ which, for the BS equation,  are the integration variables. In order
     to reduce the  four-dimensional integral  to a two-dimensional one, usually
     the phase space volume is parametrized  by the Euclidean modulus $k$,
     a hyperangle  $\chi_k$ defined  as $|{\bf k}| = k\sin\chi_k,\ k_4=k\cos\chi_k$ and by the
     standard angular space volume $d\Omega_{\bf k}$. Then the tBS vertex functions
     $\Gamma(P,k)$ are decomposed over the  hyperspherical harmonics basis which allow to carry out  the spacial
     angular integration  analytically (see for details Refs.~\cite{ourLast,ourFB,DorkinByer}).
      The resulting tBS equation represents a system of two-dimensional integral
     equations, where the vectors $p_{1,2}$ vary in  a restricted complex domain located within
     parabolas $p_{1,2}^2 = -M^2_{q\bar q}/4 +k^2 \pm iM_{q\bar q} k\cos\chi_k$.
    In  Euclidean space  the Dirac matrices    $\gamma_4=\gamma_0,\bgam_{E}=-i\bgam_{M}$ are  Hermitian and  obey
    the anti-commutation relation $\{\gamma_\mu,\gamma_\nu\}=2\delta_{\mu,\nu}$; for
    the four-product one has $(a\cdot b)={\bf a b} +a_4b_4$ .
          The masses $M_{q\bar q}$  of mesons as bound states  of a $m_1$-quark and $m_2$-antiquark follow
   from the  solution of the tBS equation at $P^2=-M_{q\bar q}^2$  in specific $J^{PC}$ channels,
   with the solution of the tDS equation (\ref{sde}) as input into the calculations in
   Eq.~(\ref{bse}). The interaction between  quarks in the  pair is encoded in  $g^2 D_{\mu\nu}$,
    imagined     as gluon exchange. For consistency, the same interaction is to be employed
     in the tDS equation~(\ref{sde}) for the inverse dressed quark propagator.

Often, the coupled equations of the quark propagator $S$, the gluon propagator $D_{\mu\nu}$ and the
quark-gluon vertex function $\Gamma_\mu$ (not to be confused with the BS vertex function
$\Gamma(P,p)$), all with full dressing (and, if needed, supplemented by ghosts and their
respective vertices), are  considered as an
integral formulation being equivalent to QCD.
  {In practice, due to  numerical problems, the finding of the  exact solution of the
system of coupled equations for  $S-D_{\mu\nu}-\Gamma_\mu$ can hardly be accomplished  and
therefore some approximations~\cite{fisher,Maris:2003vk,physRep} are appropriate.}
 Being interested in dealing with mesons as quark-antiquark bound states
 of the BS equation (\ref{bse}), one has to provide the quark
  propagator which depends on the gluon propagator and vertex as well, which in turn
  depend on the quark propagator. Leaving  a detailed discussion of the variety of approaches in dressing
 of the gluon propagator and vertex function in DS equations
 (see e.g.~Refs.~\cite{Fischer:2008uz,PenningtonUV} and references therein
 quoted) we mention only that in solving the DS equation for the quark propagator
 one usually employs truncations of the  exact interactions and replaces the gluon propagator combined with the vertex
 function by an effective interaction kernel $[g^2 D_{\mu\nu}]$.
 This leads to the tDS  equation  for
 the quark propagator which may be referred   to as the gap equation.
In explorative  calculations, the choice of the form of the effective interaction is inspired by
results from calculations of Feynman diagrams within  pQCD maintaining
requirements of symmetry and asymptotic behaviour  already implemented,
cf.~Refs.~\cite{Maris:2003vk,Roberts:2007jh,physRep,PenningtonUV}.
 The  results of such calculations, even in the simple case of accounting only for one-loop
 diagrams with proper regularization
 and renormalization procedures,  are rather cumbersome for further use in
 numerical calculations, e.g. in the framework of BS or Faddeev equations.
 Consequently, for practical purposes,  the wanted exact results are
 replaced by  suitable parametrizations of the vertex and the gluon propagator.
 Often, one employs  an  approximation which corresponds to one-loop calculations of diagrams with
 the  full  vertex function $\Gamma_\nu$,
  substituted by the free one, $\Gamma_\nu(p,k)\rightarrow\gamma_\nu$ (we suppress the color structure and
  account cumulatively for the strong coupling later on).
   To emphasize the replacement of combined gluon propagator and vertex (in the Landau gauge) we use the notation
$[g^2D_{\mu\nu}]$, where an additional power of $g$ from the second undressed vertex is
included.

\subsection{Choosing an interaction kernel}\label{choos}

 Note  that the nonperturbative behaviour of the kernel $[g^2 D_{\mu\nu}]$
 at small momenta, i.e. in the infra-red (IR) region, nowadays is
 not uniquely determined and, consequently, suitable models are needed.
There are several \textit{ans\"atze} in the literature for the IR  kernel, which can be
formally classified in the two groups: (i) the IR part is parametrized  by  two
terms - a delta distribution at zero momenta and an exponential, i.e. Gaussian term,
and (ii) only the Gaussian term is considered.
 In principle, the IR term must be supplemented
by a ultraviolet (UV) one, which assures the correct asymptotics at large momenta.
A detailed investigation~\cite{souglasInfrared,Blank:2011ha} of
 the interplay of these two terms has shown that, for bound states,
 the IR part  is dominant for light ($u$,  $d$ and $s$)  quarks
 with a decreasing role for heavier ($c$ and $b$) quark  masses  for which the
 UV part may be quite important in forming  mesons with masses $M_{q\bar q}> 3-4$ GeV as bound states.
 In the vacuum, if one is interested in an analysis of light mesons  with $M_{q\bar q}\le 3-4$ GeV, the UV term can be omitted.

Following examples in the literature~\cite{Alkofer,MT,Maris:2003vk,Krassnigg:2004if,wilson,Roberts:2007jh}
the interaction kernel in the rainbow approximation in the Landau gauge is chosen
as
\begin{eqnarray}&&
 g^2(k^2) {\cal D}_{\mu \nu} (k^2) =\left(
 D_{IR}(k^2)+ D_{UV}(k^2)\right) \left( \delta_{\mu\nu}-\frac {k_{\mu} k_{\nu}}{k^2} \right), \nonumber \\ &&
D_{IR}(k^2)=
        \frac{4\pi^2 D k^2}{\omega^6} e^{-k^2/\omega^2},
\ \quad
 D_{UV}(k^2) =
         \frac {8\pi^2 \gamma_m F(k^2)}{
            \ln\left[\tau+\left(1+\frac{k^2}{\Lambda_{QCD}^2}\right)^2\right]} ,
\label{phenvf}
\end{eqnarray}
 where the first term originates from  the effective IR part of the interaction
 determined by soft, non-perturbative effects, while the second one ensures the correct
UV asymptotic behaviour of the QCD running coupling, $ F(k^2) = \{ 1 - \exp(-k^2/[4m_t^2])\}/k^2$ with $m_t$
as an adjustable parameter, $m_t \simeq 0.5$ GeV, and  $\tau = e^2 -1$,
$\Lambda_{QCD} =0.234$~GeV, and $\gamma_m = 12/(33 - 2N_f )$ for $N_f$ as active flavors.
 In what follows we restrict ourselves to  the simplest version of the model, namely with the interaction kernel
where the UV term (the effect of which is found~\cite{souglasInfrared} to be negligibly small)
is ignored at all. As said above, this interaction
 is the  Alkofer-Watson-Weigel~\cite{Alkofer} kernel. A prominent
feature of such an interaction is that, with  only a few adjustable parameters -  $D$, $\omega$  in the IR term and
 quark mass parameter $m$ in the bare quark propagator $S_0$ - it
provides a  good description of the pseudoscalar, vector and tensor meson
 mass spectra~\cite{Hilger,Holl:2004fr,Blank:2011ha,ourLast} at zero temperature. It should be noted that, despite
the negligible contribution, the inclusion  of the UV term   in the numerical calculations causes
additional uncertainties, related to the divergency of the integrals and with requirements of regularization and
 subtraction procedures~\cite{FischerRenorm}.
 Therefore, at finite temperatures a tempting choice of the interaction  is to keep the AWW kernel the same as in vacuum.
  Notice also that, even at zero temperatures, the tBS equation
becomes rather complicate for numerical solutions since it involves the quark propagator functions
in the complex Euclidean space, where they can (actually they do) exhibit pole-like singularities. A rather detailed analysis
of solving the tBS equation in vacuum in presence of poles has been reported in Ref.~\cite{ourLast}.
Reiterate that, within the rainbow approximation, the Euclidean $P_4 = iM_{q\bar q}$ is an external parameter
 in the tBS equation, while $k_4$ is an integration variable.

 \subsection{Finite temperatures}\label{finite}
 The theoretical treatment of  systems at non-zero temperatures differs from the
 case of zero temperatures.   In this case,  a preferred  frame is determined
 by the local rest system of the  thermal bath. This means that the $O(4)$ symmetry
 is broken and,  consequently, the dependence of the quark propagator on $\bf p$ and $p_4$  requires a separate treatment.
 To describe the propagator in this case a third function $C$ is needed,
 besides the functions $A$ and $B$ introduced above for vacuum.
 { Yet, the theoretical formulation of the field theory
at finite temperatures can be performed in at least two, quite different, frameworks which
treat  fields  either with ordinary time variable
$t  \, ( - \infty  < t < \infty)$, e.g.  the termo-field dynamics
(cf.~\cite{umezawa}) and path-integral formalism (cf.~\cite{nemi,landshof}),
or with imaginary time $it= \tau$ ($0< \tau< 1/T$) which is known as the Matsubara
formalism~\cite{matzubara,kapusta,abrikosov,landsman}.
  In this paper we utilize the imaginary-time formalism within which the
partition function is defined and  all calculations are performed in Euclidean space.}
 Since, due to the  Kubo–-Martin-–Schwinger condition~\cite{periodic,KMS,Sch, kapusta}
 for equilibrium systems, the (imaginary) time evolution is restricted to the interval $[0\ldots 1/T]$,
 the quark  fields become   anti-periodic  in time with the period $1/T$.
 In such a case, the Fourier transform is not longer continuous and the energies
 $p_4$ of particles become discrete~\cite{matzubara,kapusta,abrikosov} which are known as the Matsubara frequencies, i.e.
  $p_4=\omega_n=\pi T (2n+1)$ for Fermions ($n\in \mathbb{Z}$).
 The inverse quark propagator is now parametrized as
\begin{eqnarray}
 S^{-1}({\bf p},\omega_n)=i\bgam\bp A(\bp^2,\omega_n^2)+i\gamma_4 p_4 C(\bp^2,\omega_n^2)+
B(\bp^2,\omega_n^2).
\label{inversProp}
\end{eqnarray}
Accordingly, the interaction kernel  is decomposed in to a transversal and longitudinal part
\begin{eqnarray}
[g^2 D_{\mu\nu} (\bk,\Omega_{mn})] =
\calP^T_{\mu\nu} D^T(\bk,\Omega_{mn},0)+
\calP^L_{\mu\nu} D^L(\bk,\Omega_{mn},m_g),
\label{kernel}
\end{eqnarray}
where $\Omega_{mn}=\omega_m-\omega_n$
and the gluon Debye screening    mass $m_g$ is introduced in the longitudinal part of the
propagator, where $k^2=\bk^2+\Omega_{mn}^2+m_g^2$ enters. The Debye mass
describes  perturbatively the screening of chromoelectric fields at large temperatures, therefore
it is relevant for the perturbative UV term in the limit of quark-gluon plasma where the light-quark bound states
do not longer exists, and the system ie to be described by quark and gluon degrees of greedom.
 As for the nonperturbative  pure phenomenological  IR term,  it is not a priory clear
 whether the Debye mass  has to be implemented  in the IR term or not.
 We consider both possibilities, with and without the Debye mass in the AWW model.

The scalar coefficients $D^{L,T}$ are defined below.
The  projection operators $\calP_{\mu\nu}^{L,T}$ can be written as
\begin{eqnarray}&&
\calP ^T_{\mu\nu}=\left\{  \begin{array}{llll}
                                         0,&&\mu, \nu=4,  \\
                                         \delta_{\alpha\beta}-\dfrac{k_\alpha k_\beta}{\bk^2};&&\mu,\nu=\alpha,\beta=1,2,3,
                                         \end{array}\right.
                                         \nonumber\\[1mm] &&
\calP ^L_{\mu\nu} =\delta_{\mu\nu}-\dfrac{k_\mu k_\nu}{k^2}-\calP ^T_{\mu\nu}.
\end{eqnarray}
By using the Feynman rules for finite temperatures, see e.g. Ref.~\cite{physrepMorley,kapusta}, it can be shown that
the gap equation has  the same form as in case of $T=0$, Eq.~(\ref{sde}), except that within the Matsubara formalism
the integration over $k_4$ is replaced by the summation over the corresponding  frequencies, formally
\begin{eqnarray}
\int\dfrac{d^4 k}{(2\pi)^4} \longrightarrow T\sum\limits_{n=-\infty}^{ \infty} \int\dfrac{d^3 k}{(2\pi)^3}.
\label{summation}
\end{eqnarray}
The explicit form of the system of equations for $A,B$ and $C$ to be solved can be found in
Refs.~\cite{FischerRenorm,ourModernPhys}.

The form of the  interaction kernel  is taken  the same as at $T=0$,  except that at finite $T$
the Debye mass can modify the definitions of the four momenta. The information on these
 kernels is even more sparse than in the case of $T=0$. While the effective parameters
 of the kernel in vacuum can be adjusted to some known experimental data,
 e.g. the meson mass spectrum from the BS equation, at finite temperature one
 can rely  on results of  QCD calculations, e.g. by using results of the nonperturbative lattice calculations.
 There are some indications, cf.~\cite{tereza}, that at low temperatures the gluon propagator is insensitive
  to the temperature impact, and the interaction can be chosen  as at $T=0$ with $D^T=D^L$~\cite{kitaizyRpberts}.
 However, in a hot and/or dense medium  the gluon is also subject to medium
 effects and thereby acquires modes with with
finite transversal (known also as the Meissner mass) and longitudinal (Debye or electric)
 masses. Generally, these masses  appear as independent   parameters with
 contributions depending on the considered  process~\cite{Tagaki1}.  In practice, the perturbative expressions are employed. The
role of the Meissner masses   in the tDS equation at zero chemical potential  is not yet well
established and requires separate investigations. This is beyond the goal of the present paper where
only  the role of Debye mass in  the solution of the tDS equation, $m_g$, is analysed
for the IR term. In most approaches based on the tDS equation within the rainbow approximation,
  it is also  common practice to ignore the effects of Meissner masses.
 This is inspired by the results  of a tDS equation analysis in the high temperature and density
  region~\cite{meiss} which indicate that the Meissner mass  is of no importance in tDS equation.
   At this level, the Matsubara frequencies and the Debye mass are the only $T$ depending  part of the kernel.
 The Debye mass is well defined in the weak-coupling regime. In~\cite{AlkoferDebye,Tagaki,fischerPRD90,kalinovsky}
 it was found that in the leading order
\begin{eqnarray}
m_g^2=\alpha_s
\dfrac{\pi}{3}\left[ 2 N_c+N_f\right]T^2,\label{Debye}
\end{eqnarray}
where $N_c=3$ and $N_f$ denote the number of active color and flavor degrees of freedom, respectively;
  the running coupling  $\alpha_s$ in the one-loop approximation  is
\begin{eqnarray}
\alpha_s(E)\equiv \dfrac{g^2(E)}{4\pi}=f(E) \dfrac{12\pi}{11 N_c-2N_f}
\end{eqnarray}
with $E$ being the energy scale. For the temperature range considered in the present
paper we adopt $f(E)\to 2$, which is an often employed
 choice for the Debye mass in the tDS equation~\cite{AlkoferDebye,Tagaki,fischerPRD90,kalinovsky}.
  For $N_f=4$   equation (\ref{Debye}) results in
 the  Debye mass $m_g^2 = 16\pi^2T^2/5$, which is commonly used in
  literature~\cite{roberts,kalinovsky,ourModernPhys}. It should be noted, however,  that such a choice
 of $f(E)$ is not unique. It may vary in some interval, in   dependence on
  the employed method of  infrared regularization~\cite{Tagaki,kapusta}.
  Since the Debye mass enters  as an additional energy parameter in
  $k^2=\bk^2+\Omega_{mn}^2+m_g^2$, which determines the Gaussian form of the longitudinal part of the
   AWW kernel~(\ref{phenvf}),  an increase of   $m_g^2$ results in a shift of the tDS solution towards lower temperatures
   leaving, at the same time, the shape  of the solution practically  unchanged.

The transversal and longitudinal parts of the interaction kernel~(\ref{kernel})
can be  cast in the form
\begin{eqnarray} &&
D^T(\bk,\Omega_{mn},0)=D_{IR}(\bk^2+\Omega_{mn}^2), \\ &&
D^L(\bk,\Omega_{mn},m_g)=D_{IR}(\bk^2+\Omega_{mn}^2+m_g^2).
\end{eqnarray}
In the present paper we use the parameters of the AWW model for the interaction kernel\\
  $\omega=0.5$ GeV, $D=1$ GeV$^2$, $m_l=m_{u,d}=5$ MeV, $m_s=  115$ MeV.
  Recall that the AWW model  provides  values for the vacuum quark condensate in a narrow corridor,
 $-\la q\bar q\ra_0 = [(0.244 - 0.251)$\, GeV$]^3$, and
 the correct mass spectra for pseudo-scalar, vector and tensor  mesons
  as quark-antiquark bound states~\cite{Blank:2011ha,Thomas, our2013}.

\section{Solution of the \lowercase{t}DS equation}\label{tDS}
 As mentioned above, the AWW model retains  only the IR term in the interaction kernel.
 Thus, it allows to avoid
 the logarithmic divergency  of the integrals originating from the  UV term and additional
 uncertainties related to different schemes adopted  for regularization and subtraction
 procedures,  see discussions in Refs.~\cite{FischerRenorm,Fischerrenorm1,rob-1,ourModernPhys}.
Despite the fact that,   without the UV term within the AWW model all the relevant integrals in tDS and tBS equations are convergent,  further calculations involving the solution $A,B,C$ of tDS
  can encounter  divergences (not directly connected with the interaction kernel) which have to be  properly
  regularized and  subtracted. Such a situation occurs
  in calculations of the quark condensate which, by definition,  is quadratically divergent.
For instance, at $T=0$ and $m_q\ne 0$,
\begin{eqnarray}
\la q\bar q\ra_0 =
-\frac{1}{8\pi^2}\int
\frac{  p^3 \ B(p^2) }{ p^2 A^2( p^2)+B^2( p^2)} d p,
\label{divergent}
\end{eqnarray} where $ p^2=\bp^2 +p_4^2$. At large  $p^2$
the asymptotic solution of the tDS equation becomes  $A( p^2)\to 1$, $B( p^2)\to m_q$
and the  integral (\ref{divergent}) is manifestly quadratically divergent. The same situation occurs at finite temperatures.
Usually the quark condensate is regularized by
subtracting at large momenta the asymptotic quark mass $m_q$ and defining the regularized  (subtracted)
condensate as (see,  Ref.~\cite{ourModernPhys})
\begin{eqnarray}
\la q\bar q\ra_0^l-\frac{m_l}{m_h} \la q\bar q\ra_0^h = \la q\bar q\ra_{ren.}^l-\frac{m_l}{m_h} \la q\bar q\ra_{ren.}^h,
\label{renormCondensate}
\end{eqnarray}
 where $m_l$ and $m_h$ denote the  mass of  light, e.g. $u,d$, and heavy, e.g. $s$, quarks, respectively.
  Exactly the same procedure is applied to determine  the quark condensate at finite $T$,
  see also Ref.~\cite{fischerPRD90}.
  The remaining multiplicative divergences can be removed by normalizing to
   quark condensate  at zero temperature.

\begin{figure}[!ht]
\includegraphics[scale=0.3 ,angle=0]{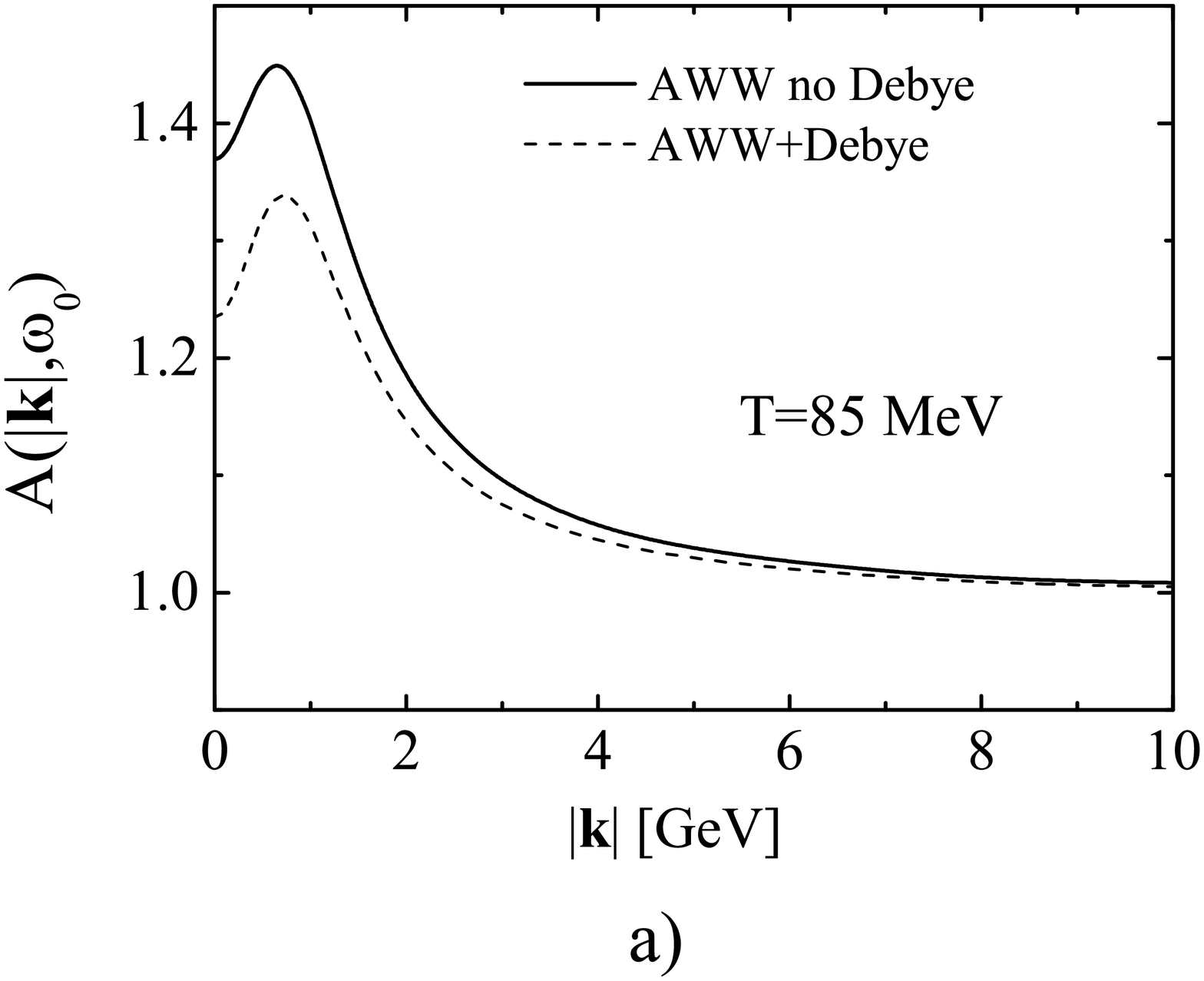}\hspace*{10mm}
\includegraphics[scale=0.3 ,angle=0]{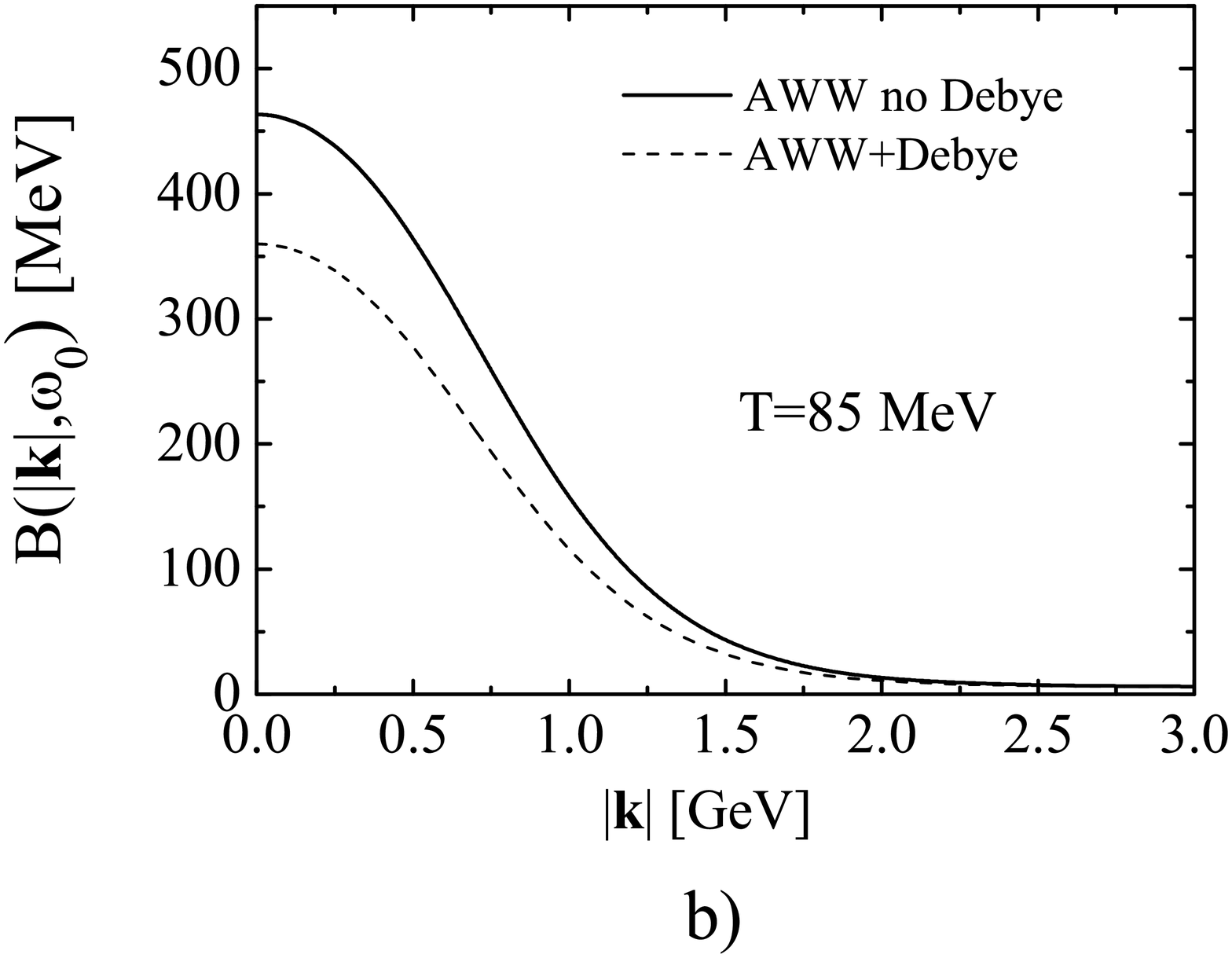}
\caption{
The solutions $A(|\bk|,\omega_0)$ (left panel) and $B(|\bk|,\omega_0)$ (right panel)
of the tDS equation for the AWW  interaction kernel
 for the light-quark mass $m_l=5$  MeV  and the lowest Matsubara frequency $\omega_0=\pi T$  at $T=85$ MeV
with  (dashed curves) and without (solid curves) Debye mass.
}
\label{fig1}
\end{figure}

We solve numerically the tDS equation for the functions $A$, $B$ and $C$  by an iteration
procedure. The summation over $\omega_n$ is truncated  at a sufficiently
large value of $n=N_{max}$, where in our calculations $N_{max}\sim 200$ for
 low temperatures and $N_{max}\sim 64$ for temperatures $T > 80 -100 $ MeV are utilized.
 The integration over the momentum $|\bk|$ is replaced by a  Gaussian integral sum,
 with the Gaussian mesh of $N_G=96$ points. In Fig.~\ref{fig1}
we present the solutions for $A$ and $B$ at the lowest Matsubara drequency $n=0$ as functions
of the three-momentum modulus $|\bf k|$ with and without taking into account the Debye mass.

  From this figure one infers that  the effect of the Debye mass is rather large at low and moderate values of the
  momentum $|{\bf k}|$. Since for the AWW kernel~(\ref{phenvf}) the Debye mass enters as a
  Gaussian exponential factor in the longitudinal part of the integral,
   i.e. $\int d|\bk|...\sim \exp(-m_g^2/\omega^2) \int d|\bk|...$,
   the effect increases with increasing  temperature, resulting in a considerable  suppression of the
  solution at large $T$. To analyse the role of the Debye mass we solved the tDS equation also in a
  large interval of $T$
  with and without the Debye mass. Notice that, since the solution of the tDS is not an observable,
 one does not have experimental quantities to be  compared with.
   Instead, one can try to reconcile   the model calculations with results of "exact" LQCD
  data. Suitable quantities of the tDS equation which can be
  related to corresponding quantities from LQCD are, e.g. the dependencies on temperature
  of the mass function $B$ and the quark condensate $\langle q\bar q\rangle$.

\begin{figure}[!ht]
\includegraphics[scale=0.3 ,angle=0]{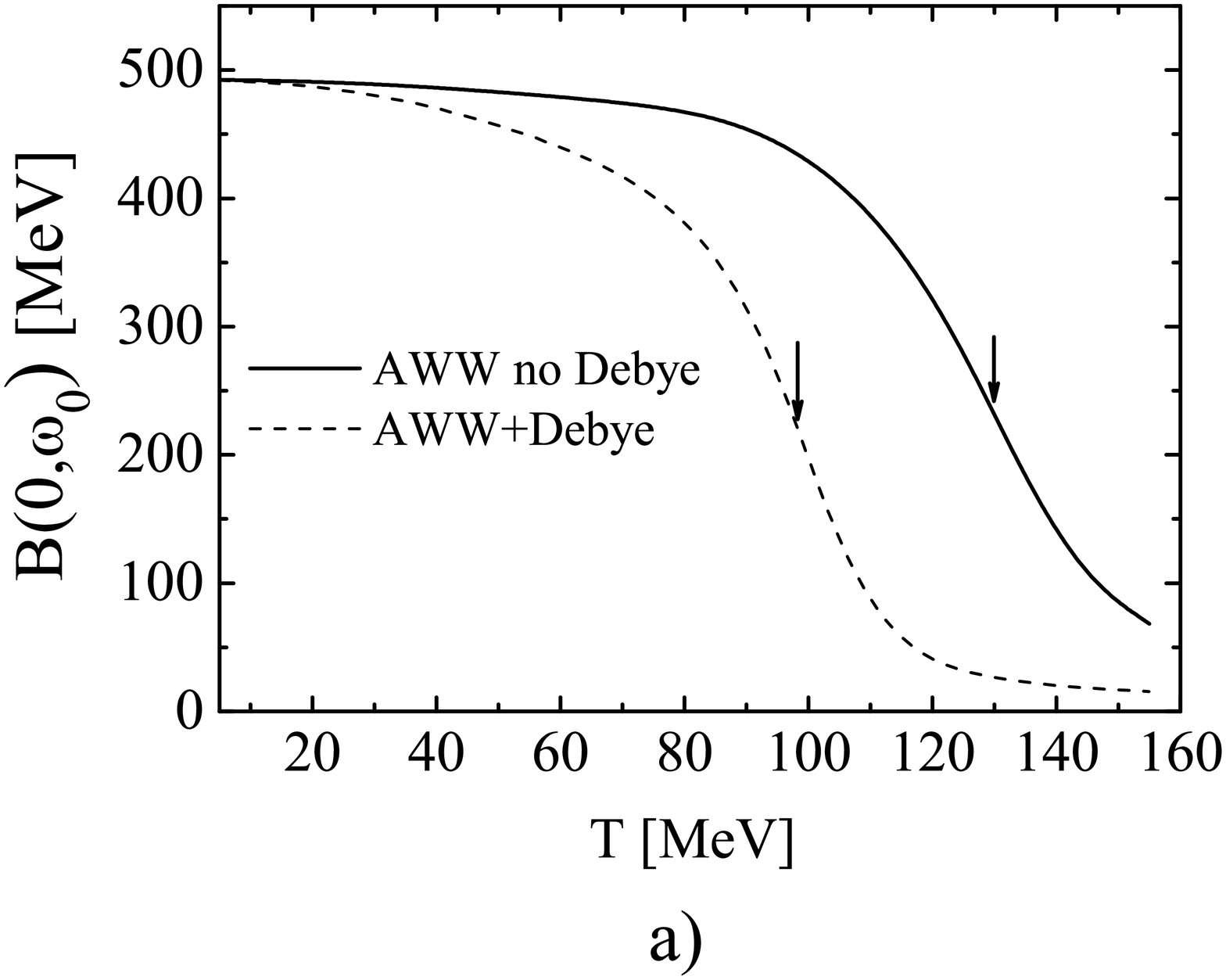}\hspace*{10mm}
\includegraphics[scale=0.3 ,angle=0]{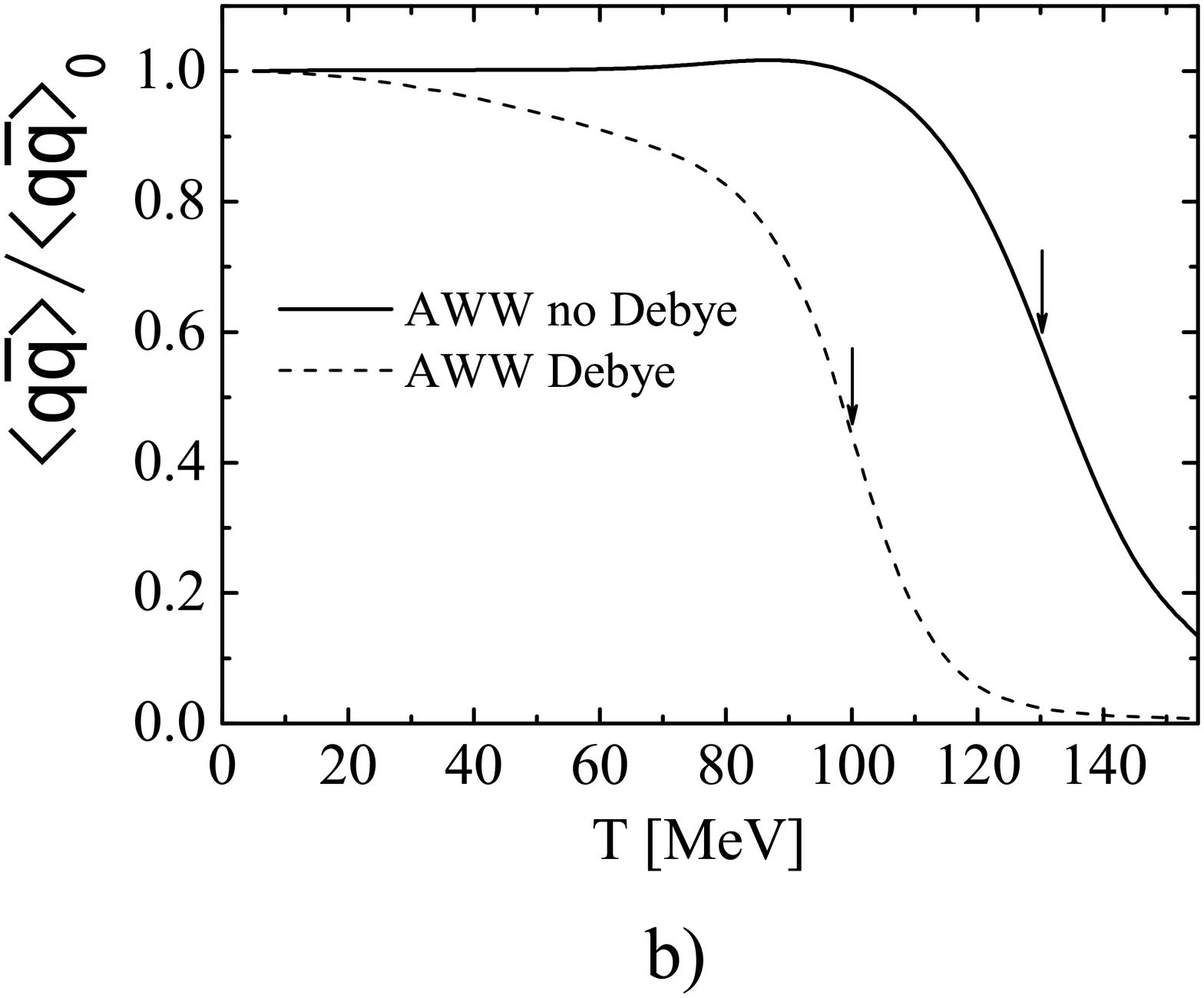}
\caption{
The solutions $B(|\bk|=0,\omega_0=\pi T)$ for the lowest Matsubara frequency $n=0$ and
 $|\bp|=0$
(left panel) and reqularized quark condensate
 $\langle q\bar q\rangle/\langle q\bar q\rangle _0$   (right panel) for the light-quark mass $m_l=5$  MeV
  as functions of temperature $T$ with
  Debye mass (dashed curves) and without the Debye mass (solid curves) taken into account. The vertical
  arrows indicate the respective position of the inflection point for the corresponding quantity.
}
\label{fig2}
\end{figure}

\begin{figure}[!ht]
\includegraphics[scale=0.3 ,angle=0]{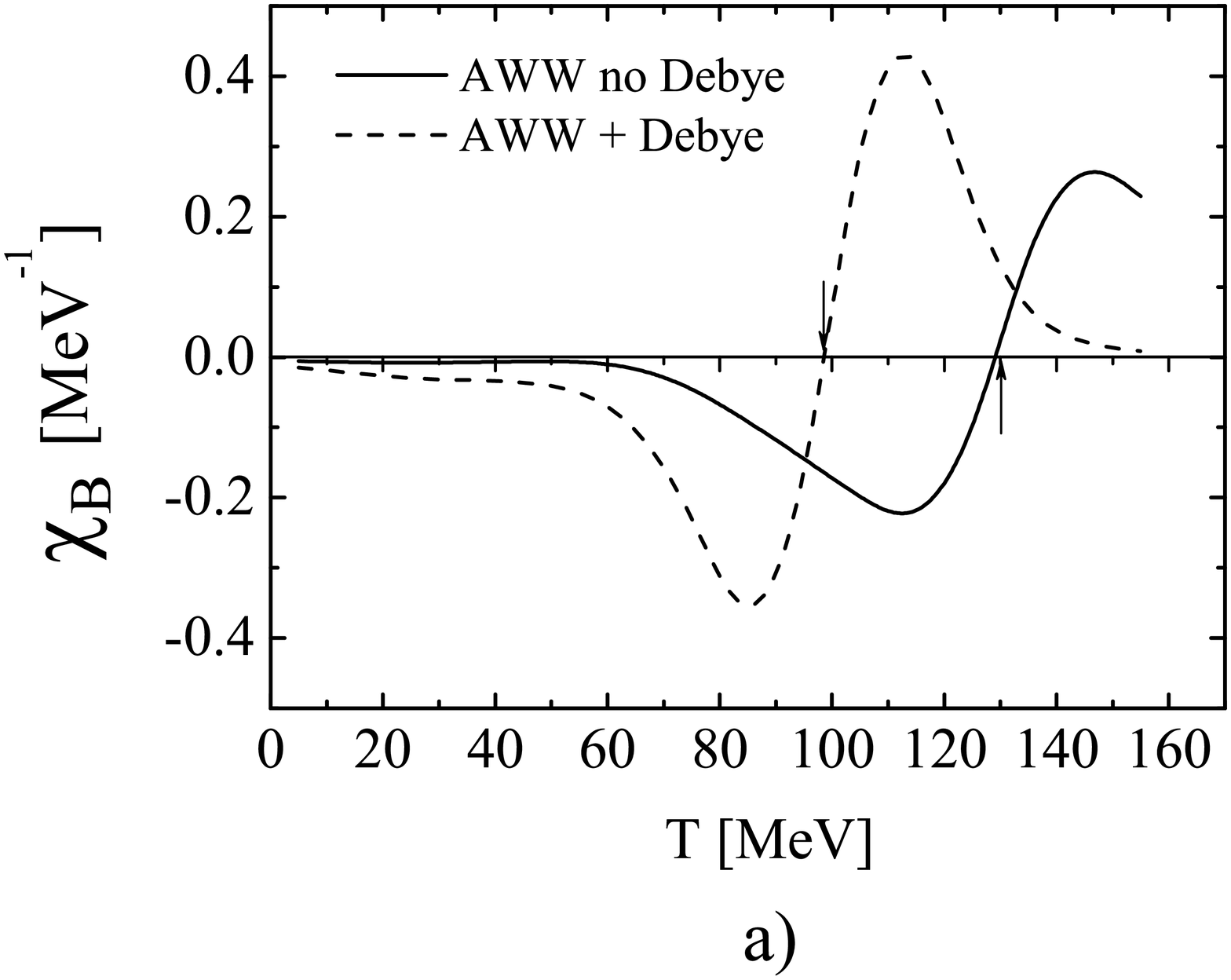}\hspace*{6mm}
\includegraphics[scale=0.3 ,angle=0]{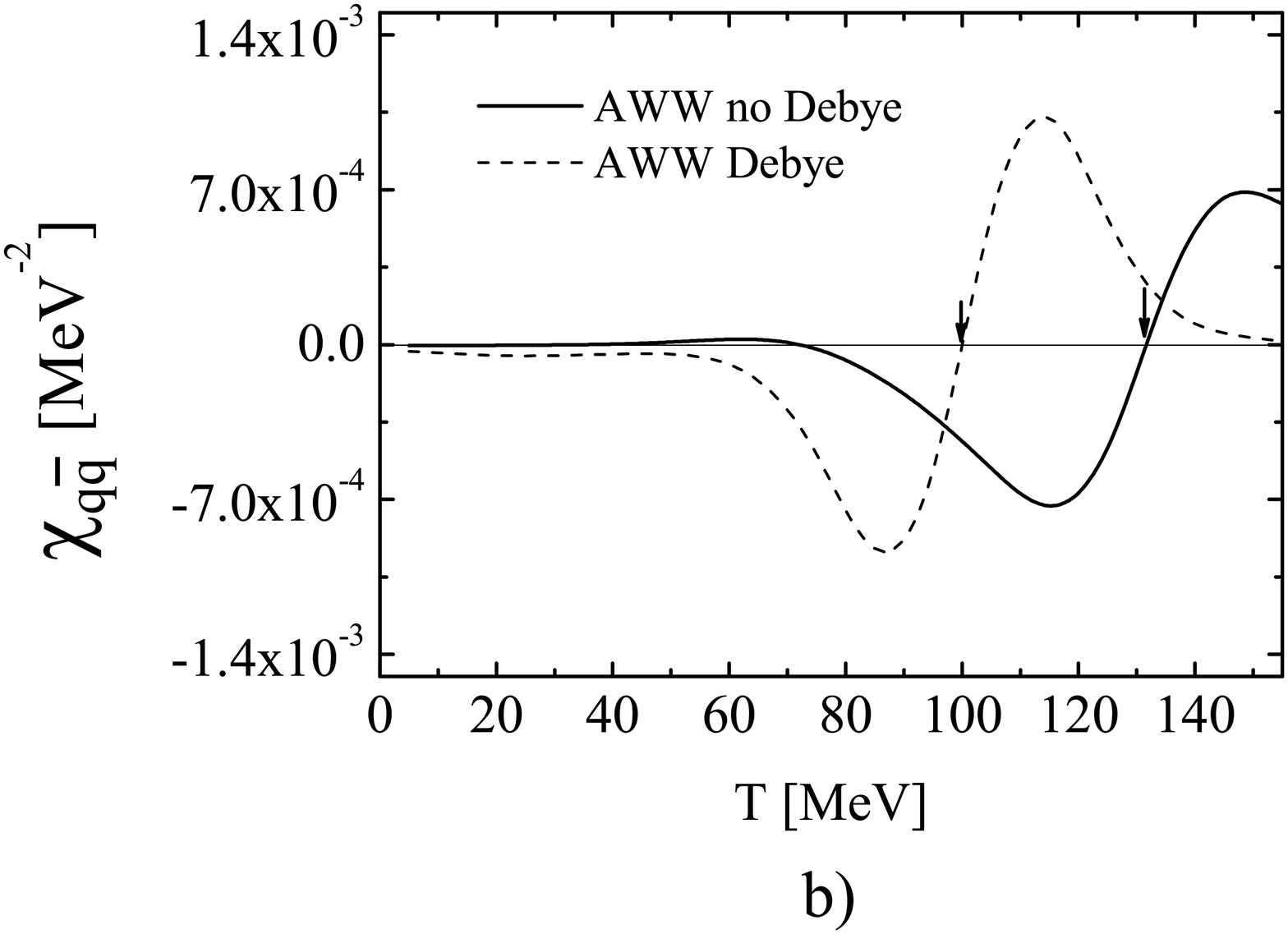}
\caption{
The inflection points (second derivative with respect to temperature)
for the mass function $B(|\bk| =0,\pi T)$ (left panel) and for the normalized quark condensate,
Eq.~(\ref{deflection}) (right panel), as exhibited in Fig.~\ref{fig2}. The arrow positions correspond
to the pseudo-critical temperatures $T_c\sim {\cal O}(100)$~MeV for the AWW with Debye mass included
and $T_c\sim {\cal O}(132)$~MeV without the Debye mass, respectively. }
\label{fig3}
\end{figure}

  Results of such calculations are presented in Fig.~\ref{fig2}, where we exhibit
  the mass function $B$, which can be considered as mass parameter according to the
  decomposition~(\ref{inversProp}), at zeroth momentum $|\bk|=0$ and zero Matsubara frequency $\omega_0 =\pi T$
  and the regularized condensate $\langle q\bar q\rangle$ (\ref{divergent}, \ref{renormCondensate})
   normalized at $T=0$.
It is seen from the figure that the solutions of the mass function  and the quark condensate, calculated
with and without the Debye mass taken into account,
 are essentially different at large temperatures. As expected,
 including the Debye mass shifts the results to lower temperatures.
To compare the model results with LQCD results
 we use the method of the maximum of the chiral susceptibility, i.e. the maxima
of the derivatives of  $B$ and/or  $\la q\bar q\ra$ with respect to the quark bare mass, as well as
the inflection point of the mass function or of the condensate, i.e. the maxima of the
corresponding derivatives with respect to the temperature~\cite{fischerPRD90}:
\begin{eqnarray}
\chi_B(T)=\dfrac{d^2 B(0,\omega_0)}{dT^2}; \ \ \chi_{q\bar q}(T)=\dfrac{d^2 \la q\bar q\ra }{dT^2}.
\label{deflection}
\end{eqnarray}
The (pseudo-) critical temperature
$ T_c$ is fixed by the condition  $ \left. \chi_B(T)\right |_{T=T_c}=0$ and/or $\left .\chi_{q\bar q}(T)\right |_{T=T_c} =0$, see Fig.~\ref{fig3}.

Figures~\ref{fig2} and~\ref{fig3} clearly demonstrate that the inflection points at finite quark bare masses
provide much smaller (pseudo-) critical temperatures $T_c$ if the Debye mass is taken into
account.  For the AWW model,  one has $T_c\sim {\cal O}(100) $ MeV, cf. also Ref.~\cite{BlankKrass}.
Without  the Debye mass in  the IR term,  the positions of the inflection points occur
  at $T_c\sim {\cal O}(130-135)$ MeV, for both, the solution $B$ and  the regularized quark condensate. This
  pseudo-critical temperature  is quite close to that obtained in LQCD calculations~\cite{yoki,FischerRenorm}
  which report $T_c\sim {\cal O}(145 - 155)$ MeV.
  This implies that, for a better agreement with the lattice results,
the model interaction kernel must not contain the Debye mass. Recall that, for the AWW model the Debye mass
enters in the longitudinal  interaction kernel exponentially, $\propto \exp(-16\pi^2 T^2/\omega^2)$, hence
sizably   suppresses the solution $B$ at large $T$.  This is a hint that, generally in models based on the rainbow
 approximation, the Debye mass has to be taken into account  only in the perturbative part of
 the interaction, e.g. in the UV term. Note that in the literature there are also other attempts to modify the
 interaction kernel  and to introduce a suitable tuned additional $T$ dependence at large $T$~\cite{kitaizyRpberts}.
 In what follows in solving the tBS equation for $q\bar q$ bound states in
 most calculations we ignore the Debye mass. However, for illustration of
  the effects of $m_g$, we present some results where the solution of the tDS contains $m_g$ as well.

 \section{\lowercase{t}BS equation at finite temperatures}\label{BSE}
 \subsection{Partial decomposition of the BS vertex function}
In the present paper we focus on the tBS equation for pseudo-scalar states.
 Prior an analysis  of  the tBS equation at finite temperature, we briefly  recall the calculations at
 $T=0$. The $O(4)$ symmetric
 solution of Eq.~(\ref{bse}) in Minkowski space for the pseudo-scalar mesons can be written in the form
 (cf. Refs.~\cite{Llewellyn,Alkofer}
 \begin{eqnarray}
\Gamma(P,p)=[{\cal F}_1 +(pP){\cal F}_2 \hat p + {\cal F}_3\hat P + {\cal F}_4(\hat p\hat P -\hat P \hat p)]\gamma_5,
\label{mink}
 \end{eqnarray}
 where the scalar vertex functions ${\cal F}_i$  are functions on $p^2$ solely. In~(\ref{mink}) the
 notation $\hat p = p^\mu\gamma_\mu$ and $\hat P = P^\mu\gamma_\mu$ is adopted.
 It has been found that the contribution of the last term in Eq.~(\ref{mink}), proportional to the tensor
 matrix $\sigma_{\mu\nu}=i[\gamma_\mu,\gamma_\nu]/2$,
 is negligibly small~\cite{blaske,ourLast} and, for the qualitative analysis we are interested in the present paper,
 can be safely omitted. Then, at $T=0$, there remain only three components in the BS vertex.

 To extend the  equation to finite temperatures one has to take into account  the
 broken $O(4)$ symmetry which  requires separate consideration of  space-like and time-like products of four vectors.
 Consequently, the first three terms in Eq.~(\ref{mink})  transform into six components:
  \be
\Gamma(P,q)\equiv \tilde \Gamma(P,q) \gamma_5=\frac12\left [
A_1 + A_2 P_0\gamma_0 - A_3 (\tilde q \tilde p) q_0 \gamma_0 -
A_4 q_0 P_0 {\hat {\tilde q}}  + A_5 (\tilde q \tilde p) {\hat{\tilde q }}- A_6{\hat{\tilde p }}
\right ] \gamma_5,
\label{gammaq}
\ee
where the scalar functions $A_i$ depend separately on $(P_0,q_0)$ and $(|\bP |,|\bq |)$.
The unit vectors $\tilde q$ and $\tilde p$ in (\ref{gammaq}) are purely spatial, i.e. $\tilde q =(0,\bq/|\bq|)$
 and $\tilde p=(0,\bP/|\bP|)$. If one considers only the zeroth Matsubara frequency for the four-vector $P$,
 $P=(0,\bP)$,  the decomposition (\ref{gammaq}) reduces to the one used before     for calculations of
 screening masses~\cite{kalinovsky}.
  To pass to Euclidean space, recall that for a meson at rest at $T=0$ the four product $(Pq)$  transforms as
  $(P_0q_0)_M \to -(P_4q_4)_E=-iM_{q\bar q}q_4$, where $q_4$ is the integration variable in the BS equation.
  Within the rainbow approximation, the interaction kernel does not depend on the total momentum $P$.
   Therefore,  in the tBS equation, $P_4$ plays the role of an external parameter
   which defines the pole of the  two-particle Green function, $P_4^2 = -M^2_{q\bar q}$. If the two-particle system
    is not at rest  then  $P_4^2 + {\bf P}^2= (iE)^2 + {\bf P}^2=-M^2_{q\bar q}$,
   where $E$ is the total energy of the meson.
  At finite temperatures the Feynman rules in Euclidean space~\cite{kapusta} result in the
  same formal procedure for the tBS equation
 as the one used in deriving the tDS equation, i.e. formally, the relative momentum becomes
 discrete, $q=(\bq, q_4)\to q_n=(\bq , \omega_n)$
 and the integration over $q_4$ is replaced by summation over the Matsubara frequencies
  $\omega_n$, cf. Eq.~(\ref{summation}). The total energy of the meson becomes also discrete,
  $iE\to iE_N=i \Omega_N=2\pi i N T$, where $\Omega_N$ is the Matsubara frequency for bosons,
   with $N \in \mathbb{Z} $.
 Then  the BS equation in Euclidean space reads
  \be
 \tilde \Gamma(P_N,p_n)   =\frac{4}{3} T\sum_m \int \frac{d^3 q}{(2\pi)^3} \gamma_\mu S^{(+)}(1) \tilde \Gamma(P_N,q_m)
 \tilde S^{(-)}(2) \gamma_\nu D_{\mu\nu} (\kappa_{mn}),
 \label{tildeBS}
 \ee
where $\tilde \Gamma$ is defined by Eq.~(\ref{gammaq}),
$\kappa_{mn}=(\bp-\bq,\omega_n-\omega_m) $ and $ \tilde S^{(-)}(2) =\gamma_5   S^{(-)}(2) \gamma_5$.
Correspondingly, the quark propagators $ S^{(+)}(1)$ and $\tilde S^{(-)}(2)$ are
\be &&
S^{(+)}(1)=i\vec\gamma\cdot \left  (\bP/2  + \bq\right )\sigma_{V}(1) -i\gamma_4\cdot
\left  (  { P_4}/{2}  +q_4\right )\sigma_{C}(1)+\sigma_{S}(1), \\ &&
 \tilde S^{(-)}(2)=i\vec\gamma\cdot\left ( {  \bP }/{2}  - \bq \right )\sigma_{V}(2)+
 i\gamma_4\cdot \left  (  { P_4}/{2}  -q_4\right) \sigma_{C}(2)+ \sigma_{S}(2),
 \label{DSEucl}
 \ee
where $\sigma_{V,C,S}$ are defined by the solution of the tDS equation at the same temperature $T$,
\begin{eqnarray}
\sigma_{V,(C,S)}(1,2)=\dfrac{A(\bq_{1,2},\omega_{m_{1,2}}), (C(\bq_{1,2},\omega_{m_{1,2}}),B(\bq_{1,2},
\omega_{m_{1,2}}))}{
\bq^2 A^2(\bq_{1,2},\omega_{m_{1,2}}) + \omega_{m_{1,2}}^2 C^2(\bq_{1,2},\omega_{m_{1,2}})+B^2(\bq_{1,2},\omega_{m_{1,2}})}.
\label{sigmas}
\end{eqnarray}

\subsection{Angular integration}
As   seen in Eqs. (\ref{tildeBS})-(\ref{sigmas})   the tBS equation
implicitly  depends on three spatial solid angles,
 ($\theta_\bP,\varphi_\bP$), ($\theta_\bq,\varphi_\bq$) and ($\theta_\bp,\varphi_\bp$).
The dependence on ($\theta_\bq,\varphi_\bq$) and ($\theta_\bp,\varphi_\bp$) follows from
 the interaction kernel $D\left(\kappa^2=(\bp-\bq)^2+(\omega_m-\omega_n)^2\right )$, which consists of two parts,
$D_1(\kappa^2)\sim {\rm e}^{-\kappa^2/\omega^2}$ and $D_2=\kappa^2/\omega^2 D_1(\kappa^2))\equiv
 \alpha\kappa^2\exp(-\alpha\kappa^2)$.
The  dependence on ($\theta_\bP,\varphi_\bP$) and  also on ($\theta_\bq,\varphi_\bq$) comes from the
propagator functions $\sigma_{F_1}(1)$ and
$\sigma_{F_2}(2)$.
 The angular parts of $D_1$ and $D_2$ of the kernel can be handled by
 decomposing them over the  spherical harmonics
 ${\rm Y}_{lm}(\bq)$ and ${\rm Y^*}_{lm}(\bp)$ as
\be
D_1(\kappa^2)\equiv{\rm e}^{-\kappa^2/\omega^2}=
{\rm e}^{-[(\omega_n - \omega_m)^2 +(|\bq| - |\bp|)^2]/\omega^2}
 4\pi  \sum_{lm} f_l^{(s)}(2|\bq| |\bp|/\omega^2) {\rm Y}_{lm}(\bq){\rm Y^*}_{lm}(\bp)
 \label{kernelD}
 \ee
where the coefficients $f_l^{(s)}(2|\bq| |\bp|/\omega^2)$ (here $\omega^2$ denotes the slope
parameter of the AWW kernel, not to be confused with a Matsubara frequency $\omega_n$) are
proportional to the scaled spherical  Bessel functions of the first kind, $I_n(x)$,
\be
f_l^{(s)}(x)=\sqrt{\frac{\pi}{2x}} {\rm e}^{-x}I_{l+\frac12}(x),
\label{fl}
\ee
with $x\equiv (2|\bq| |\bp|/\omega^2)$.
The second part of the kernel can be inferred from  Eq.~(\ref{fl}) by observing that
$D_2(\alpha \kappa^2)=-\alpha\frac{d  D_1(\alpha \kappa^2)}{d\alpha}$, where $\alpha={1}/\omega^2$.

The angular dependence on ($\theta_\bP,\varphi_\bP$) and  ($\theta_\bq,\varphi_\bq$), which comes from the
propagator functions, is more involved. Since   $\sigma_F(1,2)$ are numerical solutions of
the tDS equation, an analytical expression for the decomposition over the corresponding spherical
harmonics is lacking. Nevertheless, we decompose the product of two propagator functions $\sigma_{F_1}(1)$ and
$\sigma_{F_2}(2)$ as
\be
\sigma_{F_1}(p_1) \sigma_{F_2}(p_2)=\sum_L \sigma_L^{F_1F_2}(\omega_n,\Omega_N, |\bq|,|\bP| )
\sum_{M}      {\rm Y}_{LM}(\bP) {\rm Y}_{LM}^*(\bq) ,
\label{sigmaVV}
\ee
    where $p_{1,2}=q \pm  \frac12 P$,  and the coefficients
    $\sigma_L^{F_1F_2}(\omega_n,\Omega_N, |\bk|,|\bP| )$ have to be computed numerically
    from the solution of the tDS equation
\be
\sigma_L^{F_1F_2}(\omega_n,\Omega_N, |\bq|,|\bP| )=2\pi\int d(\cos\theta_{Pq})
\sigma_{F_1}(p_1)\sigma_{F_2}(p_2)   {\rm P}_L(\cos\theta_{Pq}).
\ee
In the  expressions above, ${\rm P}_L(\cos\theta_{Pq})$ are the Legendre polynoms;
  the superscripts  $F_1F_2$ denote the decomposition of  diagonal $F_1F_2 = VV,\ SS,\ CC$
and   non-diagonal products $VS, \ SV, \ VC,\  CV, \ CS, \ SC$ of the propagator functions, respectively.
Now we are in a position to reduce the tBS equation to a system of linear algebraic equation and   solve it
numerically.

\section{Numerical methods}\label{NumMetods}
Equation (\ref{tildeBS}) is a $4\times 4$ matrix in the Dirac spinor space.
To find the corresponding equations for the partial components $A_i (i=1\dots 6)$ we consecutively multiply the
l.h.s. and r.h.s. of the equation by the matrices
\be &&
\begin{array}{ccc}
\tilde G_1=\displaystyle\frac12 \hat I,  &\phantom{oooo}&\tilde G_2 =\displaystyle\frac12 \gamma_4,\\[2mm]
\tilde G_3 =i\displaystyle\frac12 \displaystyle\frac{\vec\gamma \bp}{|\bp|},&&
 \tilde G_4= i\displaystyle\frac12 \displaystyle\frac{\vec\gamma \bP}{|\bP|}
 \end{array}
 \ee
and compute the corresponding traces and find the integral equations of the partial
components $A_i$ as a  system of three-dimensional integral equations
with an infinite  summation over the Matsubara frequencies. Prior to the analytical   angular
integration, we solve the tDS numerically to find the coefficients
$\sigma_L^{F_1F_2}(\omega_n,\Omega_N, |\bq|,|\bP| )$ also numerically.
 Then we use the decomposition (\ref{kernelD}) for the interaction kernel and integrate
 analytically  the resulting expressions. Note  that additional angular dependencies emerge
 also from the traces which, together with (\ref{kernelD}) and (\ref{sigmaVV}), make the final expression
 rather cumbersome. It may contain products of up to four spherical harmonics for each solid
 angle of the three-vectors $\bq$, $\bp$  and  $\bP$, e.g. products of the form
 $\sum\limits_{i_i,m_i}(\cdots) {\rm Y}_{l_1m_1}(\bq){\rm Y}_{l_2m_2}(\bq){\rm Y^*}_{l_3m_3}(\bq){\rm Y^*}_{l_4m_4}(\bq)$,
 cf. Eqs.~(\ref{kernelD})-(\ref{sigmaVV}). The analytical integration results
 in a  more cumbersome expression which contains a series of products of $3j$-, $6j$- and $9j$-symbols to be
  summed up over all $l_i m_i$ for each vector $\bq$, $\bp$  and  $\bP$.
Calculations can be essentially simplified if one uses packages for analytical calculation.
 We employ the Maple packages to calculate traces and to manipulate the  Racah
symbols~\cite{racah} to perform explicitly the summation over all quantum numbers.
The remaining integration over the momentum $|\bq|$ is calculated by
discretizing the integral with a Gaussian mesh. The interval for
 $|\bq|=[0,\infty]$  is truncated by a sufficiently large value of $|\bq|_{max}={\cal O}(10-15)$~GeV.
  To re-arrange the Gaussian nodes closer
to the origin $|\bq|\sim 0$ we apply an appropriate  mapping of the  mesh by changing the variables as
\be
y=x_0 \frac{1+x}{1-x}
\ee
with $x_0$ chosen to assure $|\bq|=|\bq|_{max}$ at the last Gaussian node. In our calculations
we use  Gaussian meshes with $N_G=36$ and $N_G=48$ points, respectively.
 Eventually,  the resulting system   of linear
equations reads schematically
\begin{equation}
X= \, SX, \label{syst}
\end{equation}
where  the vector
\begin{equation}
X^T=\left (
[\{ A_1^n(q_i)\}_{i=1}^{N_G}]_{n=1}^{M_\mathrm{max}},
 [\{A_2^n( q_i)\}_{i=1}^{N_G}]_{n=1}^{M_\mathrm{max}},
 \ldots,
 [\{A_6^n( q_i)\}_{i=1}^{N_G}]_{n=1}^{M_\mathrm{max}}
\right ),
\label{vector}
\end{equation}
\noindent
 for a given Matsubara frequency $\Omega_N$ and given value of $|\bP |$,
represents the sought solution in the form of a group of sets of
partial wave components $A_\alpha^n(q_i)$,
  specified on the integration mesh of the order $N_G$ and the maximum $M_\mathrm{max}$
  for the Matsubara frequencies $\omega_n$. In our calculations we use $M_\mathrm{max}= 25 - 30$, which is
  quite sufficient for summations over $\omega_n$ within the AWW model with only the Gaussian-like IR term.
 Then the matrix $S$ is of   dimension $N_S\times N_S$, where $N_S=\alpha\times {M_\mathrm{max}}\times N_G$ with
 $\alpha=6$.
Since the system  (\ref{syst}) is homogeneous,
the  eigenvalue  solution  is obtained from the condition $\Delta=\det(S-\mathbb I)=0$.
At fixed $T$ and Matsubara frequency
$N$,  only the free external parameter remains as the three-vector of the
two-quark system $|\bP|$. The zeros of the determinant of $S-1$
are sought by changing   $|\bP|$ from a minimal value $|\bP|\sim 0$ to a maximum value $|\bP|=2\pi N T$.
More details about the numerical algorithm of solving the BS equation have been reported in
Refs.~\cite{ourLast,DorkinByer}. Note  that, with the adopted truncations for the Matsubarra  summation and
the employed   Gaussian  meshes, the resulting dimension of the determinant is rather large, $N_S\sim 5000 - 8000$.
Consequently, calculations of the  elements and the determinant itself are
lengthy and require some computer time. To increase the computer efficiency
in  solving the tDS equation and in computing the matrix elements of the matrix $S$ we use parallel computations
on a PC farm within the MPI ({\it Message Passing Interface}) standard package with a
 large enough amount of cores  ($\sim 250-300$). The efficiency  of such calculations is directly
proportional to the number of cores used in parallel.
 A more involved  situation occurs for the  determinant $\Delta$. A straightforward  use of parallel
 computation  is hindered by the fact that, if one employs  the pivoting in the Gaussian elimination method,
 the communication time among processors
increases  rapidly with an increase of number of  processors  and could be even larger than the calculations with a single
core. We developed a method of computing determinants in parallel with an optimization of the
computing and communication time in such a way that, in  our case of large dimensions,
 the total   time  can be reduced by a factor of $10-20$. The corresponding algorithm and
the employed code will be reported elsewhere.

\section{Results}
\begin{figure}[!ht]
\includegraphics[scale=0.35 ,angle=0]{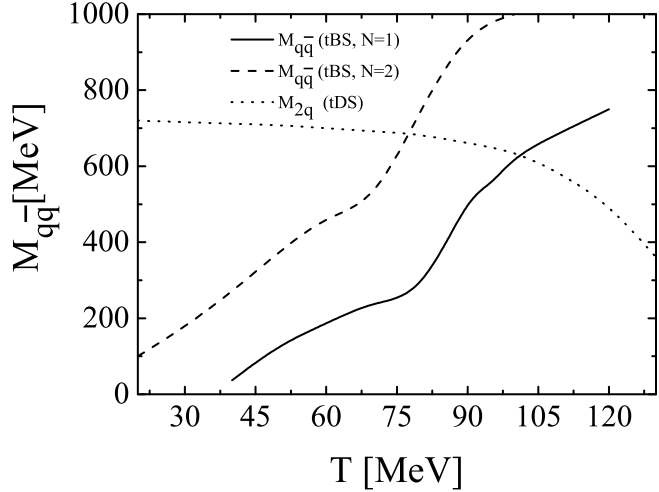}
\caption{
 The solution of the tBS equation with the AWW kernel   without the Debye mass for   pseudo-scalar mesons
 for the Matsubara frequency $N=1$, solid curve and
 $N=2$, dashed curve. The dotted line represents the maximum value of the two quasi-free quark masses as the
 solution of the tDS equation for the Matshubara frequency with  $n=0$  and zero three momentum $|\bk|=0$,
 $M_q\equiv B({\bf 0},\omega_0)/A({\bf 0},\omega_0)$, $m_u=5$~MeV and the AWW slope parameter
 $\omega^2=1$~GeV$^2$.}
\label{fig4}
\end{figure}

We solve the tBS equation for pseudo-scalar ground states for two values of the Matsubara frequency $N=1$ and $N=2$.
At   given temperature and Matsubara frequency the possible value of the pole mass is restricted
to the interval $4\pi^2 N^2 T^2 < M_{q\bar q}^2 \le 0$ which corresponds to $0<|\bP|^2 < 4\pi^2 N^2 T^2$.
The maximum value of the pole mass corresponds to the limit of thermal mass, i.e. $|\bP|\to 0$.
For the lowest Matsubara frequency, $N=1$, this limit can occur already at   $T\sim 100$~MeV,
 which implies that the solution of tBS at large temperatures approaches the thermal limit $|\bP|\to 0$.
In most of our calculations, the Debye mass has been omitted. Results of calculations for the ground state
pseudo-scalar pole masses are presented in Fig.~\ref{fig4} for the Matsubara frequencies  $N=1$ (solid curve)
and $N=2$ (dashed curve) as functions of temperature. In both cases, the mass rapidly increases with increasing
temperature, and already at $T\geq 100$~MeV (for $N=1$) and $T \geq 80$~MeV (for $N=2$)
becomes larger than the maximum value of the  mass of two quasi-free quarks at the same value of $T$.
This is demonstrated in Fig.~\ref{fig4}, where the mass of two quarks, defined as the solution of
tDS equation, $M_{2q}=2B({\bf 0},\omega_0)/A({\bf 0},\omega_0)$, is presented by the dotted curve.

For temperatures, where $M_{q\bar q} > M_{2q}$,  the bound ground  state, in the "canonical" sense, does not
exists or can be considered as unstable against the dissociation into two correlated quasi-free quarks.

\begin{figure}[!ht]
\includegraphics[scale=0.4 ,angle=0]{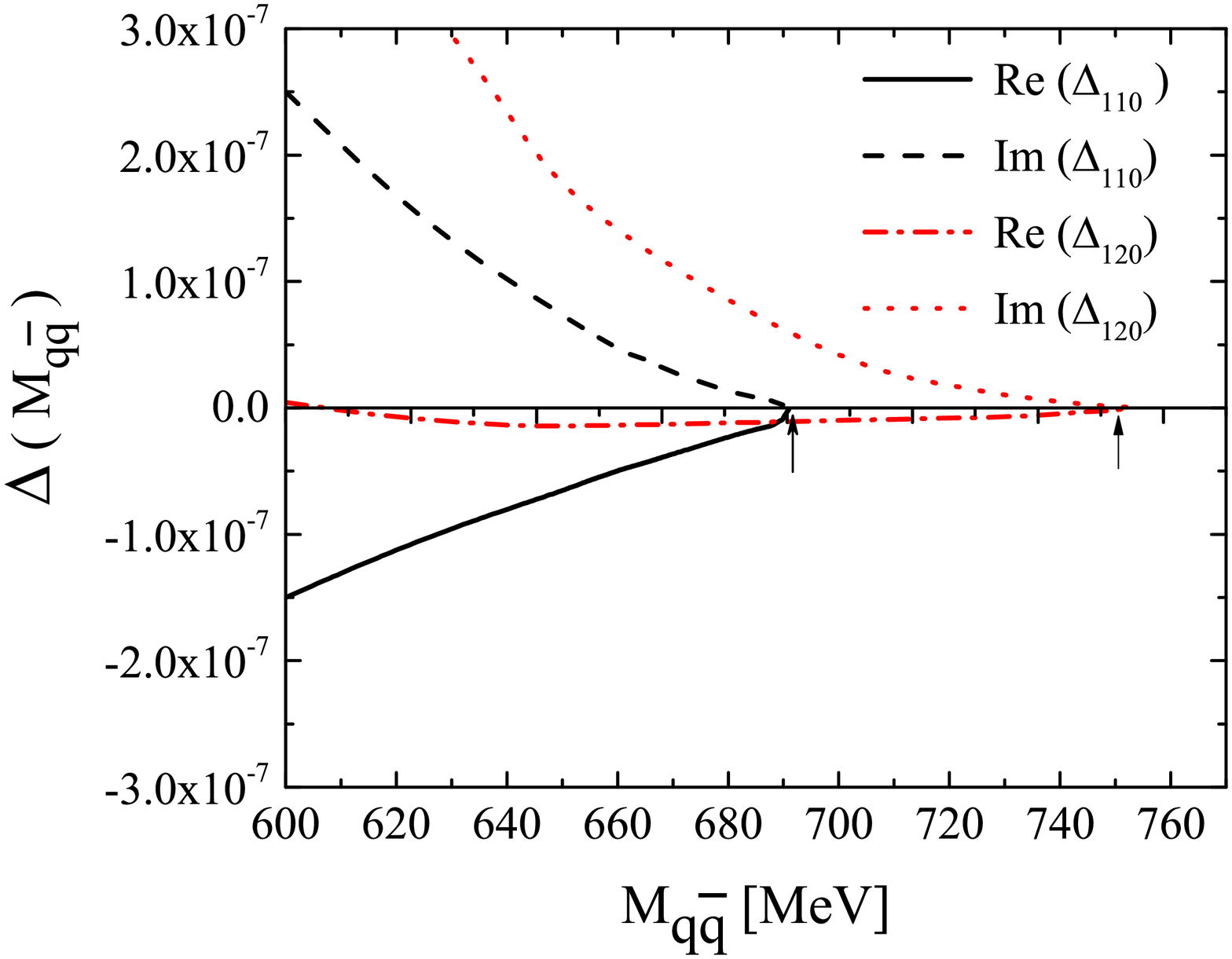}
\caption{
 An illustration of the approach of the pole mass solution to the thermal
  mass $2\pi T$ at large temperatures for the Matsubara frequency with $N=1$.
   The real and imaginary  parts of the determinant $\Delta$ at $T=110$ MeV are depicted by
   solid and dashed curves, respectively.  The real and imaginary parts at $T=120$ MeV
   correspond to dotted and dot-dashed lines. The arrows point to the limit
   $|\bf P|=0$, i.e. the thermal mass limit.}
\label{fig5}
\end{figure}

 Now we focus on the case of large temperatures for the solution with  Matsubara frequency $N=1$.
As mentioned above, the maximum value of the pole mass is limited by the
thermal mass at $\bP=0$,  i.e. $M_{q\bar q}\to 2\pi N T$.
In  Fig.~\ref{fig5}, we present an illustration of such a limit, where the real and imaginary parts
 of the determinant $\Delta$ are depicted as a function  of $M_{q\bar q}$ at two values of the temperature, $T=110$~MeV and $T=120$~MeV. It can be seen that
the determinant converges to zero, i.e. to the solution of the tBS equation, exactly in the region of the
corresponding thermal mass. In   Fig.~\ref{fig5},
  this is depicted by  two vertical arrows. Note that, at large values of
$T$  and for $N=1$, we have not found  solutions of the  tBS equation  with smaller pole masses.

\begin{figure}[!ht]
\includegraphics[scale=0.4 ,angle=0]{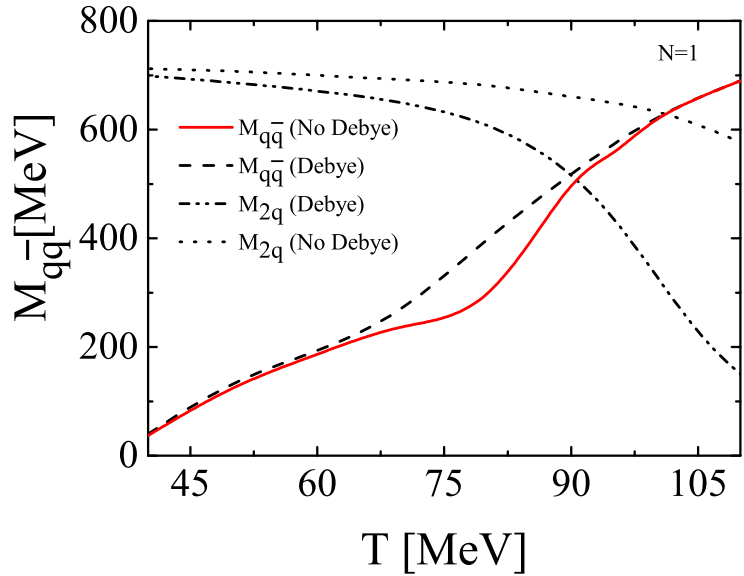}
\caption{
 Comparison of the tBS solution for the Matsubara frequency with $N=1$ with (dashed curve)
  and without (solid curve) taking into account
 the Debye mass in the tDS equation.  The   dash-dotted  and dotted curves represent the sum of
 masses $M_q\equiv B({\bf 0},\omega_0)/A({\bf 0},\omega_0)$ of two free quarks obtained from
 the solution of the tDS equation, also with and without the Debye mass, respectively.}
\label{fig6}
\end{figure}

 We investigate now also the influence  the Debye mass on the bound states for $N=1$. For this
we consider the tBS equation with the solution of tDS equation with Debye mass taken into account.
 Results are exhibited in Fig.~\ref{fig6}, where the solid curve denotes the pole mass without the Debye mass,
 cf. Fig.~\ref{fig4}, and the dashed curve illustrates the effect of the Debye mass.
 As expected, at low temperatures,
 where $m_g$ is small, the two solutions are practically the same. At intermediate values of $T$,  the interaction
 kernel is more compact, hence  it leads to larger values of the masses. At larger temperatures
 both solutions approach the thermal mass limit, see Fig.~\ref{fig5}.
 For completeness, in Fig.~\ref{fig6} we also present the values of two quasi-free quark masses
 $M_{2q}\equiv 2B({\bf 0},\omega_0)/A({\bf 0},\omega_0)$, as solutions of tDS equation with
 Debye mass (dot-dashed curve) and without Debye mass (dotted curve).
  It is seen that with Debye mass taken into account the
 pole mass becomes larger than the two free masses at already $T\sim 90$~MeV, i.e.
 the dissociation occurs at lower temperatures. This is in
 agreement with the behaviour of the inflection points $\chi_B$
  of the mass solution $B$   and $\chi_{q\bar q}$ of the quark condensate $\la q\bar q\ra$, cf.  Fig.~\ref{fig3}.

\section{Summary}\label{summary}
 We have investigated   the solution of the truncated Dyson-Schwinger  (tDS)
 equation  at finite temperature within the rainbow approximation by employing
 the Alkofer-Watson-Weigel (AWW) model which consists  only
 of the infra-red term in a more complex interaction kernel. The  solution
 of the tDS equation is a prerequisite for a
 consistent solution of the truncated Bethe-Salpeter (tBS) equation for
 quark-antiquark bound states at finite temperature within the
 same approximation.   The ultimate goal is to use  the tBS equations,
  for an analysis of the behaviour of hadrons in hot matter, including possible (phase) transitions
 and  dissociation effects. For this goal we investigate to what extent the model, which provides a fairly good
 description of ground-state mesons at zero temperatures, can be applied to the truncated tDS equation at finite temperatures.  We find that a direct inclusion of the Debye mass  in the interaction kernel results
  in a too low value of the pseudo-critical temperature $T_c$, defined as the inflection point of the mass function
  $B({\bf 0},\omega_0)$ or/and of the normalized quark condensate $\la q\bar q\ra$, being by ${\cal O}(50\%) $ smaller than  the ones found in lattice QCD calculations. Since in the considered model at a given temperature   $T$,  the  Debye mass  $m_g$ enters into the longitudinal part of the interaction
  as a (constant) Gaussian exponential factor $\propto\exp(-16\pi^2 T^2/5\omega^2)$,
   it makes problematic the attempts of obtaining larger values  of the QCD deconfinement temperature
   $T_c$ close to  the lattice values.
   We argue that, since the Debye mass $m_g$ is essentially a perturbative quantity, it
  has to be taken into account in the perturbative ultra-violet term to maintain a correct behaviour
 at large momenta, while in the nonperturbative infra-red term it can be omitted at all. We find that by
  omitting the Debye mass $m_g$  one is able to obtain critical temperatures closer
  (smaller by only 10-15\%) to that from lattice $2+1$ flavour QCD.   To achieve a better agreement with lattice data the  IR term requires further meticulous analyse of the tDS equation at finite
  temperatures  which will be done elsewhere.

  As in the case of zero temperature, $T=0$, we consider the tBS equation within the rainbow approximation with the quark propagators
  entering the tBS equation found preliminarily as solutions of the tDS equation at the same
  temperature.  In our calculations, we restrict the boson Matsubara frequency $\Omega_N=2\pi N T$ to $N=1$ and $N=2$. Larger values
  of $N$ provide  large values of $M_{q \bar q}$ ($N>2$), far above masses of known lightest pseudo-scalar mesons.  For each Matsubara frequency $\Omega_N$,   the ground state  mass $M_{q\bar q}^2 = 4\pi^2 N^2 T^2-\bP^2$
   is defined as the first (lowest) zero of the corresponding determinant as a function of $|\bP|$. Such  a mass is referred to as  the Matsubara pole mass.

  For both frequencies  $\Omega_N$, $N=1$ and $N=2$, we find that the pole
  masses rapidly increase with increase of  temperature. This is in agreement with the behaviour
  of the screening masses at large temperatures reported in literature, see e.g.
  Refs.~\cite{yaponetz,kalinovsky}.
It has been found that, for the lowest Matsubara frequency, where  $N=1$,  a solution of the tBS
equation at large temperatures, $T>110$~MeV, can be found only in  the limit of thermal masses, i.e. at
  $|\bP| \to 0$.  We analyse  the behaviour of the pole masses as a function of temperature
  with respect to the $T$ dependence of the masses of free quarks obtained from the
  tDS equation.
  At large values of $T$, $T>100$ MeV,  the  pole masses become larger than the sum of two  quarks.
  This implies that at large $T$ the ground state of two quark does not
  occur in the sense   as commonly adopted  in quantum mechanics where the binding energy is negative.
  This can be interpreted as dissociation instability of the state
  against a fragmentation  into a state  of two quasi-free quarks.
   The Matsubara pole masses do not have a direct relation to the inertial masses, however, one can
   expect that the obtained behaviour of the pole masses follow the same tendency as the
   inertial masses at the same temperature.

\section*{Acknowledgments}
This work was supported in part by the Heisenberg - Landau program
of the JINR - FRG collaboration.  LPK appreciates the warm hospitality at the
Helmholtz Centre Dresden-Rossendorf. The authors acknowledge discussions with Dr. J. Vorberger.

\end{document}